\documentclass[amsmath,amssymb,twocolumn,aps,prb,floatfix]{revtex4-1}
\usepackage{comment}
\usepackage{graphicx,ams math,epsfig,epstopdf}
\usepackage{dcolumn}
\usepackage{bm}
\usepackage{tabularx}
\usepackage{color}
\usepackage{mathrsfs}
\usepackage{ulem}
\usepackage{hyperref}
\hypersetup{
	colorlinks = True,
	urlcolor=blue,
	linkcolor = blue,
	citecolor= blue
}

\begin{document}

\title{Learning Thermodynamics with Boltzmann Machines}

\author{Giacomo Torlai and Roger G. Melko}

\address{
Perimeter Institute for Theoretical Physics, Waterloo, Ontario N2L 2Y5, Canada\\
Department of Physics and Astronomy, University of Waterloo, Ontario N2L 3G1, Canada
}

\date{\today}

\begin{abstract}
A Boltzmann machine is a stochastic neural network that has been extensively used in the layers of deep architectures
for modern machine learning applications.
In this paper, we develop a Boltzmann machine that is capable of modelling thermodynamic observables for physical systems 
in thermal equilibrium.
Through unsupervised learning, we train the Boltzmann machine on data sets constructed with spin 
configurations importance-sampled from the partition function of an Ising Hamiltonian 
at different temperatures using Monte Carlo (MC) methods. 
The trained Boltzmann machine is then used to generate spin states, for which we compare thermodynamic observables 
to those computed by direct MC sampling.  We demonstrate that the Boltzmann machine can faithfully 
reproduce the observables of the physical system.
Further, we observe that the number of neurons required to obtain 
accurate results increases as the system is brought close to criticality. 

\end{abstract}

\pacs{}%

\maketitle

\section{Introduction}

Machine learning is a paradigm whereby computer algorithms are designed to learn from -- and make predictions on -- data. 
The success of such algorithms in the area of classifying and extracting features from large data sets
relies on their ability to infer them without explicit guidance from a human programmer.  
Such automatic encoding proceeds by first ``training'' the algorithm on a large data set and then asking the trained machine 
to perform some task. 
Currently, many machine learning applications are performed with neural networks, which essentially fit the data to a graph structure composed of many nodes and edges. If the ultimate goal is to perform classification, like in image or speech recognition, the network can be trained on a labelled data set by maximizing the output probability of the correct label ({\it supervised} learning).
However, since labelled data is often scarce, a more effective strategy is to learn the full distribution of the data using a generative model, which does not require labels 
({\it unsupervised} learning). Such generative training allows the network to extract more information, and also to generate approximate samples of the distribution. 
For the classification of data, this training is followed by a supervised fine-tuning, 
which can be done with only a small amount of labelled data.

Although neural networks have been researched for many decades, the performance 
required for solving highly complex problems in real-world applications has been achieved only relatively recently with deep learning.\cite{LeCun15}  
Here, the networks are made up of several layers stacked such that the output of one layer becomes the 
input of the next layer. The ability to learn multiple levels of representations makes deep learning a very powerful tool in capturing 
features in high-dimensional data,\cite{Hinton07} and it drastically improved the performance in complex tasks such image recognition,\cite{Krizhevsky12} speech recognition\cite{Hinton12} or natural language understanding.\cite{Collobert11} Machine learning also has many applications in physics, and has 
been successfully used to {solve complex problems, including searching for exotic particles in high-energy physics,\cite{Baldi14}, solving dynamical mean-field theory in strong correlated systems \cite{Arsenault15}
or classifying the liquid-glass transition.\cite{Ceriotti16} More recently, neural networks has been also employed to 
identify phases of matter with and without conventional order parameters,\cite{Carrasquilla16} and locate the position of phase transitions to high accuracy.\cite{Wang16}
In light of this success, one may ask whether neural networks can be trained 
for other difficult problems, such as reproducing statistical-mechanical distributions 
of classical Hamiltonians in an unsupervised setting.  This would allow one, for example, to train a 
neural network using data that has been importance-sampled using  Monte Carlo (MC) from a partition function, and then to calculate estimators from the 
distribution produced by the neural network.

A natural candidate neural network for this task is a Boltzmann machine.  A Boltzmann machine is a stochastic neural network, composed of neuron-like nodes forming a network with undirected edges.  Each neuron has a binary value
that has a probabilistic element, which depends on the neighbouring units to which it is connected.
The connecting edges weigh inputs to each neuron to define its state.
This architecture, once elaborated, can be used to produce approximate reconstructions of the original data set.  More precisely,
a reconstruction is an estimate of the probability distribution of the original input, which is of course imperfectly
contained in the limited-size training data set.  
This procedure has been widely successful, leading Boltzmann machines to become a core piece
of deep learning architectures. 

In this paper, we explore the ability of Boltzmann machines to learn finite-temperature distributions of the 
classical Ising Hamiltonian and, consequently, associated thermodynamic observables such as energy, magnetization, or specific heat. We show that faithful recreation of observables is possible for a finite-size lattice Ising system.  
We also demonstrate that the number of neurons in the networks required to recreate data at the
 critical point can be much larger than in the paramagnetic or ferromagnetic phase.  This suggests that deep networks may 
be required for the faithful representation of thermodynamics by Boltzmann machines at critical points.\cite{Mehta14}

\section{The Boltzmann Machine}

In constructing a Boltzmann machine, our goal is to build an approximate model of a target probability distribution.  For the sake of 
concreteness, we will consider the Boltzmann distribution of $N$ Ising spin variables, weighted by the partition function, as our
target distribution.  It is natural to imagine sampling this distribution with a MC procedure.  In addition to producing these 
samples, a MC simulation usually calculates estimators of thermodynamic observables, such as energy or specific heat,
directly from the sampled target distribution.
However, one could instead imagine obtaining estimators from an approximate distribution constructed to mimic our target distribution.  
In this scenario, spin configurations can be generated by a Boltzmann machine that was trained by the MC samples
of the target distribution.  In this section, we review the concept of sampling the target distribution for an Ising spin Hamiltonian, and
detail the construction and training of a Boltzmann machine.   In Sec.~\ref{sec:results} we present the results for thermodynamic
observables obtained from this Boltzmann machine, trained on finite-temperature configurations produced from the nearest-neighbor Ising ferromagnet.

\subsection{Target probability distribution and thermodynamic observables}

Consider a system of $N$ classical spins on a $d$-dimensional lattice, with Ising spin configuration
${\bm {\sigma}} = \{\sigma_1, \sigma_2, \cdots, \sigma_N \}$, and
a generic  Hamiltonian $H_S(\bm{\sigma})$
where the $S$ subscript indicates the physical (spin) system. When the system is at thermal equilibrium at temperature $T$, the ``target'' probability of a spin configuration $\bm{\sigma}$ is given by the familiar Boltzmann distribution
\begin{equation}
p_S(\bm{\sigma},T) = \frac{1}{Z_S} \text{e}^{-H_S(\bm{\sigma})/T} 
\end{equation}
where $Z_S = \text{Tr}_{\bm{\sigma}}\text{e}^{-H_S(\bm{\sigma})/T} $ is the canonical partition function. With the knowledge of $Z_S$
it is possible to compute all thermodynamic potentials and average values of observables. However, the estimation of the partition 
function involves a summation over all the $2^N$ states, which is feasible only for very small systems. The average value of an
observable $\mathcal{O}$ can be calculated as
\begin{equation}
\langle\mathcal{O}(T)\rangle = \frac{1}{M}\sum_{k=1}^M\,\mathcal{O}(\bm{\sigma}_k) \label{avgO}
\end{equation}
if $\bm{\sigma}_k$ are samples drawn from the distribution $p_S(\bm{\sigma},T)$ at temperature $T$. 
This equation is exact only when $M \rightarrow \infty$.  However,
the sampling process can be done using Markov Chain MC simulations, leading Eq.~\eqref{avgO}
to give an expression for a MC expectation value for finite but large $M$.  
In the below, we consider expectation values obtained with this procedure to be the 
exact results for the target probability distribution.  They will be compared to observables 
calculated from a probability distribution generated by a Boltzmann machine, as we now describe.




\subsection{Restricted Boltzmann Machine}
\begin{figure}[t]
\includegraphics[width=80mm]{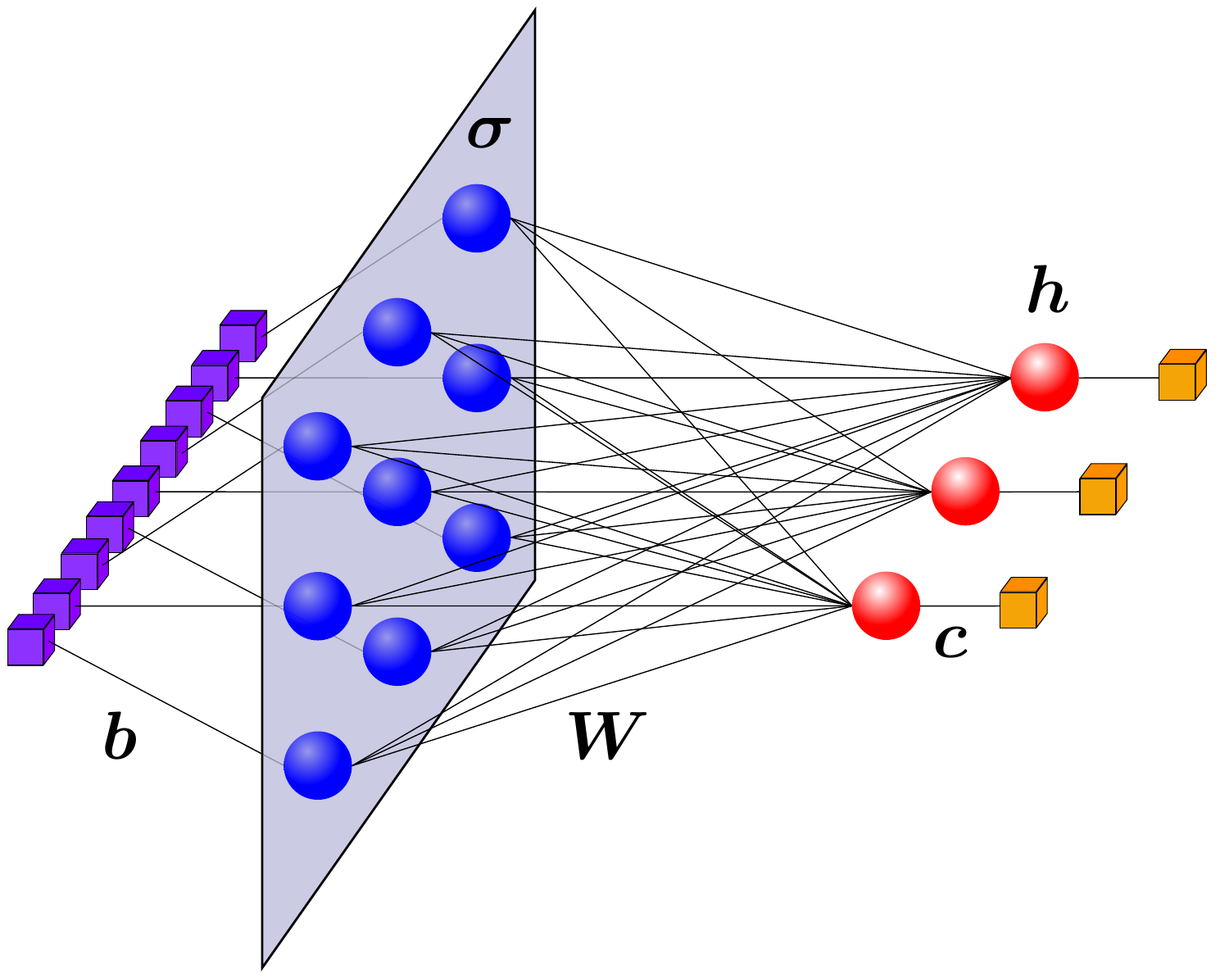}
\caption{Restricted Boltzmann machine. The visible units (blue) are connected to the hidden nodes (red) with a symmetric matrix of weight $\bm{W}$. The external fields in the Hamiltonian are represented by new edges with weights $\bm{b}$ and $\bm{c}$ connecting the visible and hidden nodes respectively with ancillary units (purple and orange) with value clamped to one.}
\label{rbm}
\end{figure}
Given a target probability distribution $p_S(\bm{\sigma})$ defined over a set of random variables $\bm{\sigma}$, our goal is to build a probabilistic model $p_{\bm{\lambda}}(\bm{\sigma})$ which mimics our target distribution. The model is in general characterized by a set of parameters $\bm{\lambda}$, which we will tune in order to minimize the distance between these two probability distributions. It is advantageous to build a joint probability distribution on a graph, where conditional independence between random variables in the corresponding probabilistic model can be better understood with the help of graph theory and through visualization. We recall that a graph is a collection of nodes and edges where to each node is associated a variable $\sigma$ and each edge represents a probabilistic relation between nodes. A probabilistic graphical model defines a joint probability distribution $p_{\bm{\lambda}}(\bm{\sigma})$ over the graph and conditional independence between the variables $\sigma$ provides us with a factorization rule for the distribution. 
We build the probability distribution over an undirected graph satisfying a local Markov property (called a Markov random field).
In particular, we adopt a bilayer architecture. 
Symmetric edges connect spin nodes $\bm{\sigma}\in\{0,1\}^{N}$ in the so-called ``visible'' layer, with ``hidden'' nodes $\bm{h}\in\{0,1\}^{n_H}$ in the hidden layer (Fig.~\ref{rbm}). The weights of the edges are described by a matrix $\bm{W}$ with zero diagonal, where the element $W_{ij}$ is the weight on the edge connecting $h_i$ to $\sigma_j$. We also introduce two external fields $\bm{b}$ and $\bm{c}$ coupled to the visible and hidden layers respectively.  One can consider the latter as weights on new edges between each visible and hidden nodes and an ancillary node, with its variable ``clamped'' (or fixed) to one. Moreover, all the variables in the graph are stochastic, comprising one major difference between this model, called a restricted Boltzmann machine, and regular neural networks. The full probability distribution defined by the graph can be written as a Boltzmann distribution
\begin{equation}
p_{\bm{\lambda}}(\bm{\sigma},\bm{h}) = \frac{1}{Z_{\bm{\lambda}}}\,\text{e}^{-E_{\bm{\lambda}}(\bm{\sigma},\bm{h})}
\end{equation}
where the model parameters are $\bm{\lambda} =\{\bm{W},\bm{b},\bm{c}\}$ and the energy is given by
\begin{equation}
E_{\bm{\lambda}}(\bm{\sigma},\bm{h})=-\sum_{ij}\,W_{ij}h_i\sigma_j-\sum_j\,b_j\sigma_j-\sum_i\,c_ih_i .
\end{equation}
As now the joint distribution is defined over two sets of nodes, the graph distribution over the spins is obtained by marginalization
\begin{equation}
p_{\bm{\lambda}}(\bm{\sigma}) = \sum_{\bm{h}}p_{\bm{\lambda}}(\bm{\sigma},\bm{h})=\frac{1}{Z_{\bm{\lambda}}}\,\text{e}^{-\mathcal{E}_{\bm{\lambda}}(\bm{\sigma})}
\end{equation}
where we introduced an effective visible energy
\begin{equation}
\label{free_energy}
\mathcal{E}_{\bm{\lambda}}(\bm{\sigma})=-\sum_j\,b_j\sigma_j -\sum_i\,\log(1+\text{e}^{\,c_i+\sum_j\,W_{ij}\sigma_j}),
\end{equation}
often called the ``free energy'' in literature on restricted Boltzmann machines.   
This probabilistic graphical model has a very important property used 
in the inference process of the states of the two layers. Since the state of any node is sampled from a non-linear 
function of its inputs  (its ``activation''), and the activations of nodes in the same layer are independent from each other (Fig.~\ref{rbm}), it is possible to sample one layer at a time, exploiting fast linear algebra routines in numerical simulations.
Moreover, for a specific choice of $\bm{\lambda}$, the states of visible and hidden layers can be inferred exactly with the posteriors 
$p_{\bm{\lambda}}(\bm{\sigma}\,|\,\bm{h})$ and $p_{\bm{\lambda}}(\bm{h}\,|\,\bm{\sigma})$.
Because the Boltzmann machine is restricted (meaning no intra-layer connections), the posteriors factorizes nicely as
\begin{eqnarray}
\label{conditionalProb1}
p_{\bm{\lambda}}(\bm{\sigma}\,|\,\bm{h})=\prod_j\,p_{\bm{\lambda}}(\sigma_j\,|\,\bm{h}), \\
p_{\bm{\lambda}}(\bm{h}\,|\,\bm{\sigma})=\prod_i\,p_{\bm{\lambda}}(h_i\,|\,\bm{\sigma}).
\label{conditionalProb2}
\end{eqnarray}
All the probabilities can be easily estimated using Bayes theorem
\begin{equation}
p_{\bm{\lambda}}(\sigma_j=1\,|\,\bm{h})=
\sigma\big(\sum_i\,W_{ij}h_i+b_j\big)
\label{conditional}
\end{equation}
with the function $\sigma(x)=(1+\text{e}^{-x})^{-1}$ called a ``sigmoid'' (a similar expression is obtained for the conditional of the hidden layer).
We point out that, although we are interested here in the generation of visible spin states, it is straightforward to extend this network for discriminative tasks. By adding a new layer for the labels, the resulting three-layer neural network can perform classification with competitive accuracies on common benchmarks.\cite{Louradour11,Larochelle08,Schmah08,Larochelle12}
Restricted Boltzmann machines also play a central role in deep learning, for instance in the greedy layer-by-layer pre-training of deep belief networks~\cite{Hinton06,Salakhutdinov08} or in their natural multilayer extension called deep Boltzmann machine.~\cite{Salakhutdinov09,Salakhutdinov12}

\subsection{Training}
We have discussed how the Boltzmann machine can generate an arbitrary probability distribution, provided a large enough number of 
hidden nodes, and how we can obtain the probability $p_{\bm{\lambda}}(\bm{\sigma})$. As we already mentioned, the training process 
consists of tuning the machine parameters $\bm{\lambda}$ until the $p_{\bm{\lambda}}(\bm{\sigma})$  is close to the target distribution 
$p_S(\bm{\sigma}$). This is equivalent to solving an optimization problem where the function to minimize is the distance between the two 
distributions.  This distance can be defined by the Kullbach-Leibler (KL) divergence 
\begin{equation}
\mathbb{KL}\,(p_S\,||\,p_{\bm{\lambda}}) \equiv \sum_{\bm{\sigma}}\,p_S(\bm{\sigma})\log \frac{p_S(\bm{\sigma})}{p_{\bm{\lambda}}(\bm{\sigma})}\ge 0
\end{equation}
with equality only if the two distributions are identical. 
We build  a 
data set $\mathcal{D}=\{\bm{\sigma}^{(1)},\dots,\bm{\sigma}^{(|\mathcal{D}|)}\}$ by drawing samples $\bm{\sigma}$ from 
the Ising
$p_S(\bm{\sigma})$ with Markov chain MC sampling at temperature $T$. The probability distribution underlying the data set is 
$p_{\text{data}}(\bm{\sigma})=\frac{1}{|\mathcal{D}|}\sum_{\bm{\sigma}^\prime}\,\delta(\bm{\sigma},\bm{\sigma}^\prime)$ and, if the 
data set size $|\mathcal{D}|$ is large enough, $p_{\text{data}}(\bm{\sigma})$ is then a good approximation of $p_S(\bm{\sigma})$. We 
can then write the KL divergence as
\begin{equation}
\mathbb{KL}\,(p_{\text{data}}\,||p_{\bm{\lambda}})= -\frac{1}{|\mathcal{D}|}\sum_{\bm{\sigma}\in\mathcal{D}}\,\log p_{\bm{\lambda}}(\bm{\sigma}) - \mathbb{H}(p_{\text{data}})
\end{equation}
where the first term is called negative log-likelihood and $\mathbb{H}(p_{\text{data}}) = -\sum_{\bm{\sigma}}\,p_{\text{data}}(\bm{\sigma})
\log p_{\text{data}}(\bm{\sigma})$ is the entropy of the data set. The optimization problem is solved by stochastic gradient 
descent. We choose an initial point $\bm{\lambda}^{(0)}$ in the full configuration space with zero external fields and weights $W_{ij}$ randomly drawn from a uniform distribution centered around zero. Gradient descent optimization consists of updating all the parameters with the rule
\begin{equation}
\bm{\lambda}_j \leftarrow\bm{\lambda}_j - \eta \nabla_{\bm{\lambda}_j}\mathbb{KL}\,(p_{\text{data}}\,||\,p_{\bm{\lambda}}).
\label{update}
\end{equation}
The size $\eta$ of the gradient step, called the ``learning rate'', is kept constant during the training. The increments in the parameters are obtained 
by averaging the gradient of the KL divergence over the entire data set $\mathcal{D}$. 
However, since the data set is usually redundant, the updates can be evaluated on a mini-batch of samples instead, resulting in a larger number of updates for each data set sweep. This optimization procedure, called stochastic gradient descent, substantially speeds up the learning, especially when the data set contains a very large number of samples. 
On the other hand, for data sets with moderate number of samples, a common issue in the training of neural networks is overfitting the training data set. Different techniques have been 
proposed to regularize the networks and overcome the overfitting, such as introducing a weight decay term in the KL divergence cost 
function,~\cite{Krogh92} or randomly removing some hidden nodes in the network (called ``dropout''~\cite{Srivastava14}).
However,  producing training data is not an issue for the cases studied here,
where MC sampling is fast and efficient.  Thus, 
we build a data set sufficiently large to avoid using regularization.
However, one could envision other cases where MC samples are expensive, so that regularization would be required.

To obtain an update rule for the gradient descent we need to take the derivative of the KL divergence in Eq.~(\ref{update}), which reduces to the 
derivative of the log-likelihood, 
\begin{equation}
\label{nabla_LL}
\nabla_{\bm{\lambda}_j}\log p_{\bm{\lambda}}(\bm{\sigma})=-\nabla_{\bm{\lambda}_j}\mathcal{E}_{\bm{\lambda}}(\bm{\sigma})+ \sum_{\bm{\sigma}} p_{\bm{\lambda}}(\bm{\sigma})\nabla_{\bm{\lambda}_j}\mathcal{E}_{\bm{\lambda}}(\bm{\sigma}).
\end{equation}
If we consider for instance the case of  $\lambda = \bm{W}$, the derivative of the visible energy is 
\begin{equation}
\label{nabla_epsilon}
\nabla_{\bm{W}}\,\mathcal{E}_{\bm{\lambda}}(\bm{\sigma})=-\sum_{\bm{h}}\,p_{\bm{\lambda}}(\bm{h}\,|\,\bm{\sigma})\,\bm{\sigma}\,\bm{h}^\top.
\end{equation}
Plugging this back into Eq.~(\ref{nabla_LL}), we obtain
\begin{equation}
\nabla_{\bm{W}}\mathbb{KL}\,(p_{\text{data}}\,||\,p_{\bm{\lambda}})=-\langle \bm{\sigma}\,\bm{h}^\top\rangle_{p_{\bm{\lambda}}(\bm{h}\,|\,\bm{\sigma})} +\langle \bm{\sigma}\,\bm{h}^\top\rangle_{p_{\bm{\lambda}}(\bm{\sigma},\bm{h})}.
\label{nablaKL}
\end{equation}
The first average of the correlation matrix $\bm{\sigma}\,\bm{h}^\top$ can be easily computed by clamping the spin variables $\bm{\sigma}$ 
to the sample from the data set, and inferring the state $\bm{h}$ of the hidden variables from the conditional distribution 
$p_{\bm{\lambda}}(\bm{h}\,|\,\bm{\sigma})$. In the second term however, the correlation matrix is averaged over the full model distribution 
$p_{\bm{\lambda}}(\bm{\sigma},\bm{h})$, which involves knowledge of the partition function $Z_{\bm{\lambda}}$. To overcome this 
issue, we instead run a MC for $k$ Markov steps 
\begin{equation}
\bm{\sigma}^{(0)}\rightarrow\bm{h}^{(0)}\rightarrow\bm{\sigma}^{(1)}\rightarrow\bm{h}^{(1)}\rightarrow\dots\rightarrow\bm{\sigma}^{(k)}\rightarrow\bm{h}^{(k)} 
\end{equation}
by sampling each layer using the exact conditional distributions. 
The updates of the stochastic gradient descent are then obtained by taking the average of Eq.~(\ref{nablaKL}) over a mini-batch $\mathcal{D}^{[b]}$ of samples
\begin{equation}
\bm{\lambda}_j \leftarrow\bm{\lambda}_j - \frac{\eta}{|\mathcal{D}^{[b]}|} \sum_{\bm{\sigma}\in\mathcal{D}^{[b]}} \nabla_{\bm{\lambda}_j}\mathbb{KL}\,(p_{\text{data}}\,||\,p_{\bm{\lambda}}).
\end{equation}
with $b=1,\dots,|\mathcal{D}|/|\mathcal{D}^{[b]}|$.
This training algorithm is called contrastive divergence
\cite{Hinton02} (CD$_k$) and is the most effective known tool for the training of restricted Boltzmann machines. Note that since the initial 
state of the chain is a sample from the data set and thus it already belongs to the distribution, there is no need for a long equilibration time.  Hence the order $k$ of the chain can be very low, 
resulting into a very fast learning procedure. In some cases, only one step (CD$_1$) is sufficient to reconstruct the visible states with low 
error. 

\section{Results} \label{sec:results}

The classical spin system we choose to train the Boltzmann machine on is the Ising Hamiltonian,
\begin{equation}
H_{S}(\bm{\sigma}) = -J\sum_{\langle ij\rangle}\,\sigma_i\sigma_j, \label{IsingHAM}
\end{equation}
with ferromagnetic interactions $J=1$ between nearest neighbours. As an instructive example we begin by training one machine on a one-dimensional 
chain with 6 spins. For such a small system it is possible to compute the partition function, and thus the full probability distribution, exactly.
We prepare a  data set of configurations using the exact probability distribution and then train a Boltzmann machine 
using CD$_5$. Because the partition function of the Boltzmann machine is known, we can compute the KL divergence for various sets $
\bm{\lambda}$, evaluating the performance of the training. By plotting the KL divergence as a function of the training steps 
(Fig.~\ref{1d}a) we see how the distribution generated by the machine improves towards the data set distribution. We also show the 
comparison between the true probability distribution and the ones produced by the machine at two different stages of the training 
(Fig.~\ref{1d}b).
\begin{figure}[t]
\centering
\includegraphics[width=80mm]{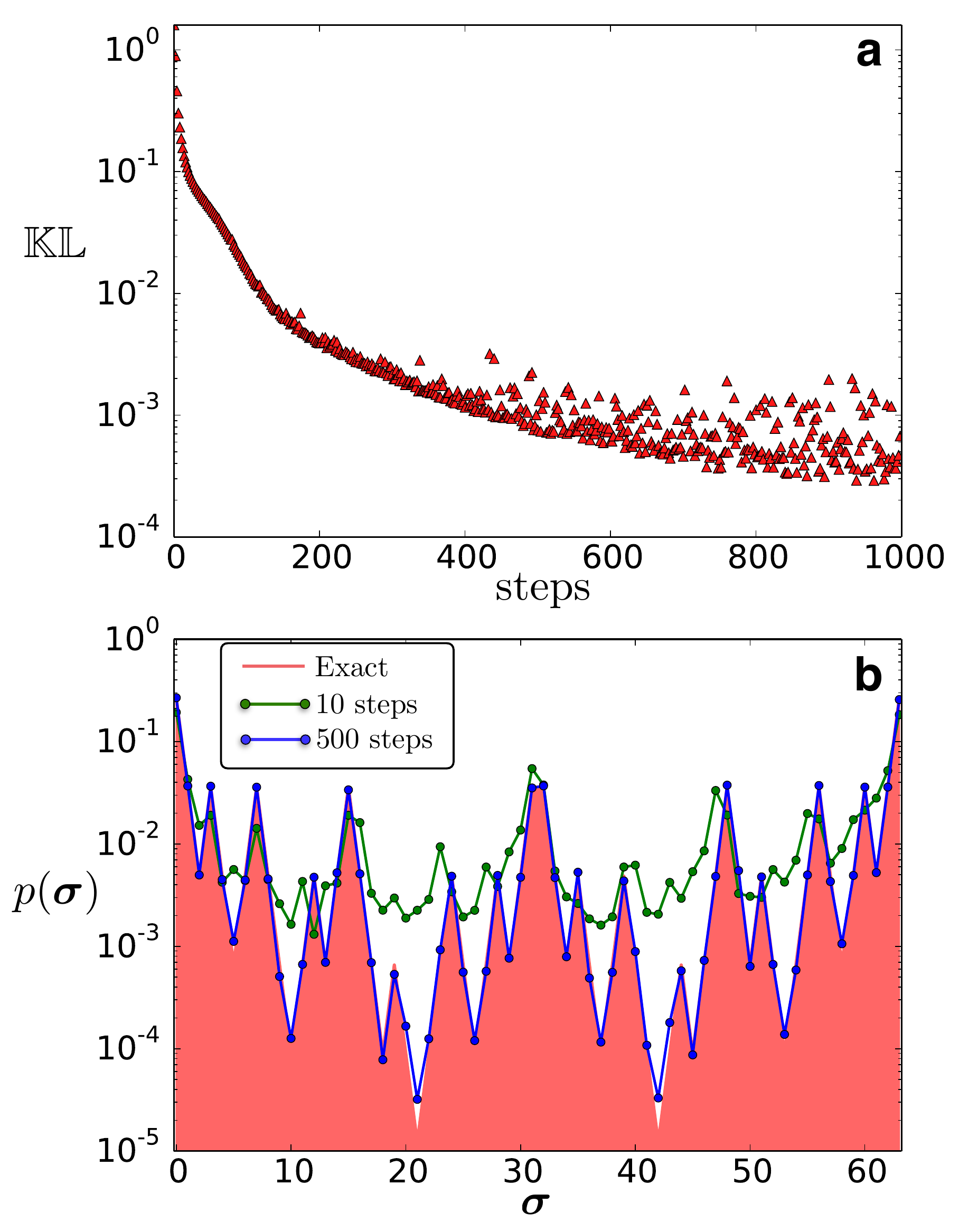}
\caption{KL divergence as a function of training step ({\bf a}) and probability distributions ({\bf b}) for a $d=1$ Ising model with $N=6$ 
spins. We show the comparison between the exact probability distribution (red) and the approximate distribution produced by the Boltzmann machine 
after 10 (green) and 500 (blue) training steps for all of the $2^6$ states $\bm{\sigma}$.}
\label{1d}
\end{figure}

Next, we consider the more interesting case of a two-dimensional system with $N=L \times L$ spins on a square lattice with periodic boundaries. Contrary to 
the one-dimensional case, this system undergoes a second order phase transition at $T_c\simeq2.269$ from an ordered ferromagnetic phase ($T<T_c
$) to a disordered paramagnetic phase ($T>T_c$).
We prepare a data set $\mathcal{D}_T$ with $10^5$ binary spin configurations MC-sampled from $p_S(\bm{\sigma},T)$ for several temperatures in a range 
centered around $T_c$.  For each $T$ we train a different machine $\mathcal{M}_\tau$ which generates a distribution $p_{\bm{\lambda}_\tau}
(\bm{\sigma})$, where the subscript $\tau$ refers to the physical temperature $T$. 
For each machine we collect samples using a different number of hidden nodes while adopting the same external hyper-parameters (learning rate, mini-batch size, number of training steps, initial conditions, etc.). 
We update the 
parameters with CD$_{20}$ using stochastic gradient descent with learning rate $\eta=0.01$ and mini-batch size of 50 
samples. We initialize the weights $\bm{W}$ from an uniform distribution around zero and width $w\propto\sqrt{1/(n_H+N)}$. 
We note that, although a larger value of contrastive divergence order $k$
is bound to improve the learning, it also substantially slows down the time required to reach a solution.
\begin{figure}[t]
\centering
\includegraphics[width=80mm]{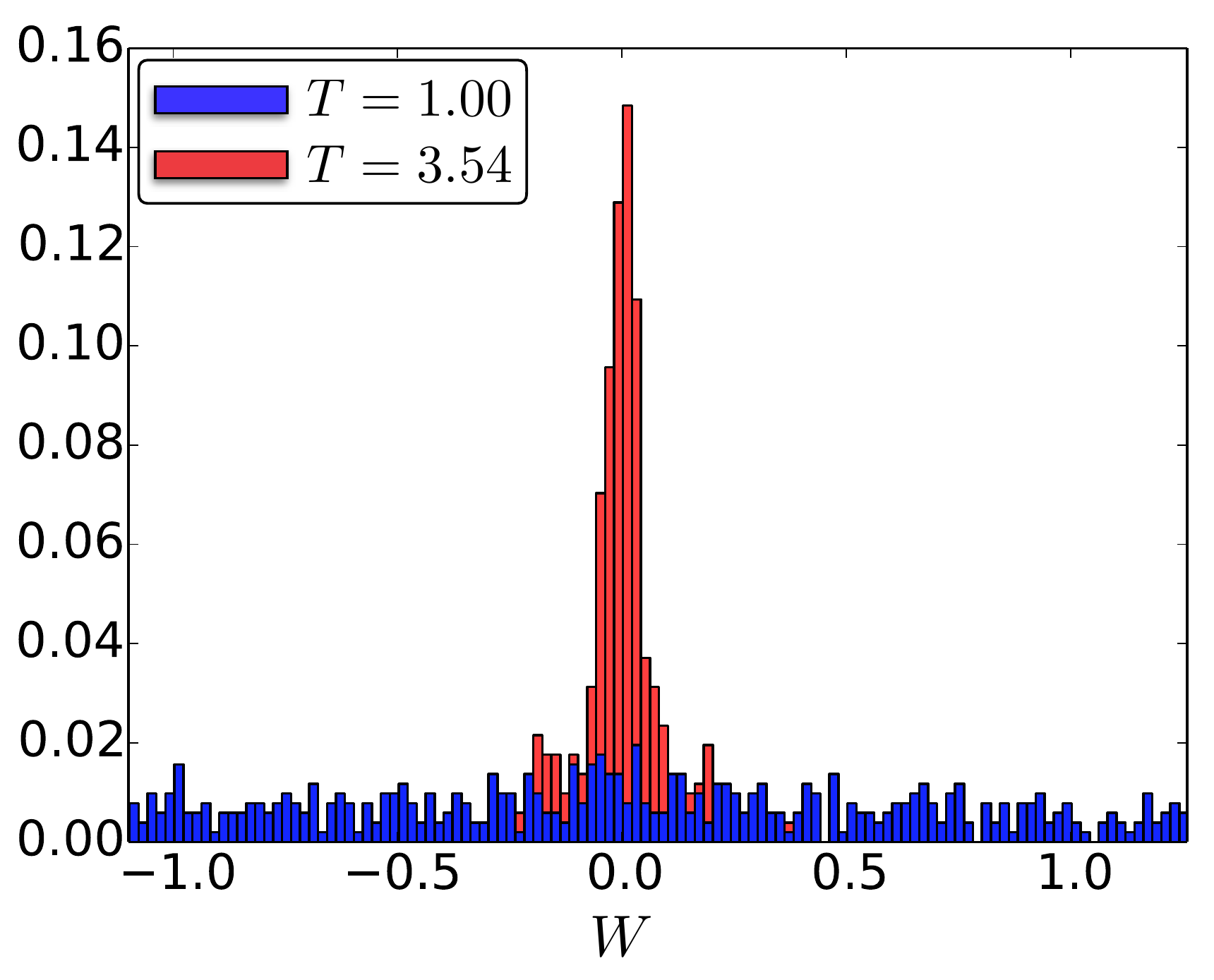}
\caption{Histogram of the relative frequency of appearance of the weight amplitudes for two Boltzmann machines with $n_h=32$ hidden nodes, trained at low and high $T$ for the $d=2$ Ising model with $N=64$ spins.}
\label{histogram}
\end{figure}
\begin{figure*}[ht]
\centering
\includegraphics[width=135mm]{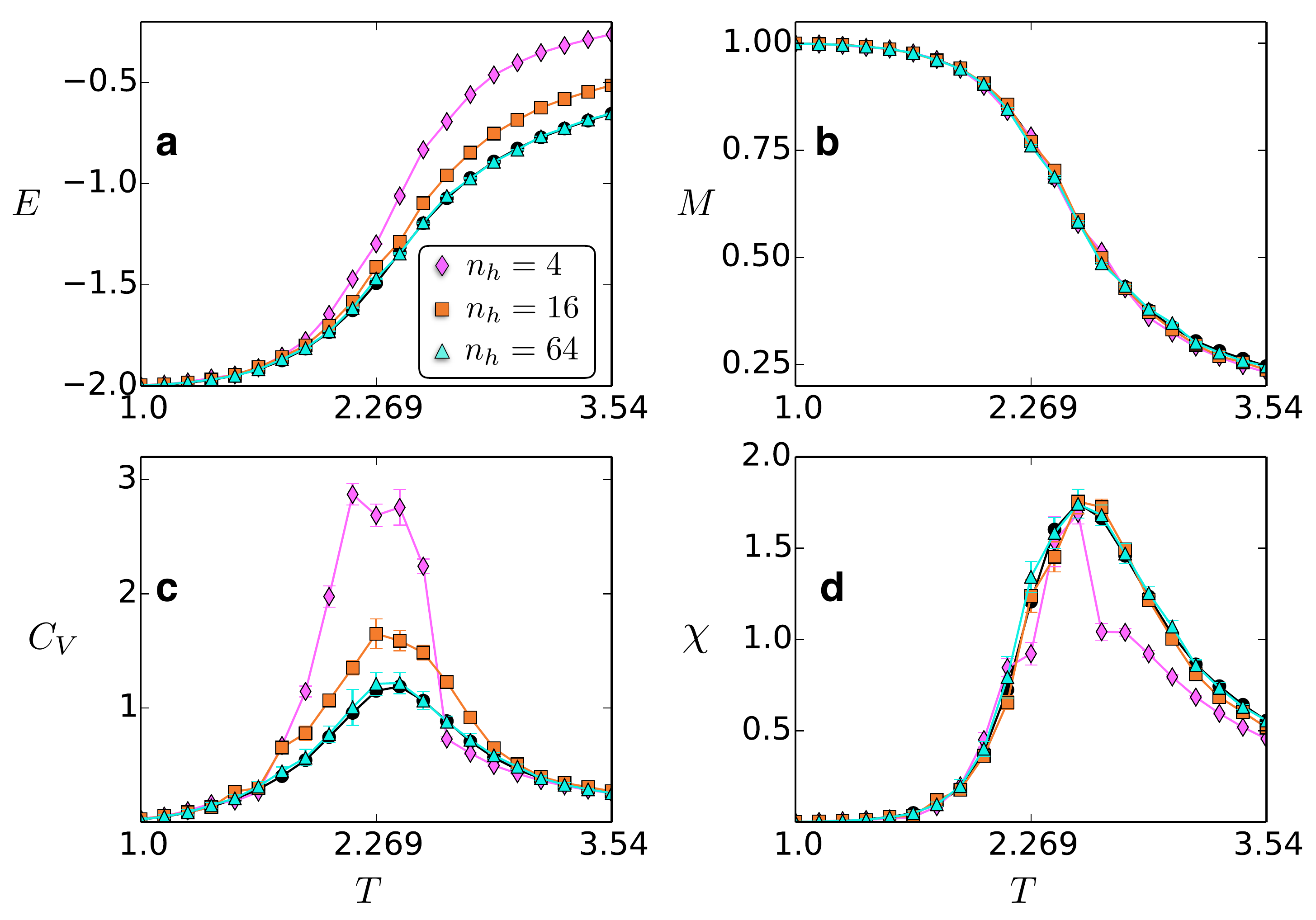}
\caption{Comparison of the observables generated with the Boltzmann machine with the exact values calculated from the data set 
(black) for a $d=2$ Ising system with $N=64$ spins. The observables considered are energy ({\bf a}), magnetization ({\bf b}), 
specific heat ({\bf c}) and magnetic susceptibility ({\bf d}). We show the results for Boltzmann machines with  
hidden nodes $n_H=4$  (pink), $n_H=16$ (orange) and $n_H=64$ (cyan).}
\label{observables}
\end{figure*}

It is natural to ask how the performance of each Boltzmann 
machine is affected when the training samples are taken at high or low temperature. Moreover, we are interested in whether or not a 
Boltzmann machine is able to properly capture the fluctuations that the system undergoes at criticality. Before discussing the quantitative analysis 
of the thermodynamics, we give an insight into the functioning of these machines by showing the histogram of the matrix elements of $
\bm{W}$ (Fig.~\ref{histogram}) after the training at low and high temperature. In these two limits we know what the probability distribution 
$p_S(\bm{\sigma},T)$ looks like and we can thus obtain a qualitative understanding of the training and sampling processes of the 
machines. At very high temperature $J/T\ll1$ the spins are completely random, so $p_S(\bm{\sigma})\simeq N/2$. In this case the 
weights histogram of the high temperature machine ($T=3.54$) displays a sharp peak
centered around zero. This means that the visible and hidden layers 
are quasi-decoupled, and the visible state is random since the activation probability from Eq.~(\ref{conditional}) is 
$p_{\bm{\lambda}}(\sigma_j\,|\,\bm{h})\simeq 1/2$. On the other hand, at low temperature the two polarized states $\bm{\sigma}=\bm{0},
\bm{1}$ are most probable and this causes the histogram to be wide and flat. When we start the sampling we initialize both visible and 
hidden layers randomly. There is a spontaneous symmetry breaking and the machine chooses one of the two polarizations. If the machine 
chooses the visible state $\bm{\sigma}=\bm{1}$ after equilibration, we find, by inspecting the hidden states driving the spins, that the hidden layer 
is arranged such that only the nodes connected to the positive weights are active (and similarly for the opposite state). The activations will be in 
this case large and positive  and thus $p_{\bm{\lambda}}(\bm{\sigma}=\bm{1}\,|\,\bm{h})\simeq 1$. Note that, even though the data set is 
completely ergodic, once the visible layer has equilibrated into one polarization state, it is unlikely to switch to the 
other. This ergodicity issue is analogous to that faced by local Metropolis updates in MC simulations of the low-temperature
ferromagnet.

We turn now to discuss performance on the full range of temperatures. 
Since for our system it is very challenging to compute the partition function and thus the KL divergence,
we instead characterize the performance of the machine using Ising thermodynamics observables. Given an observable $
\mathcal{O}$ defined on the spin system we can compare its average value computed on the spins in the dataset at temperature $T$,
\begin{equation}
\langle \mathcal{O}(T)\rangle_{\mathcal{D}} = \frac{1}{|\mathcal{D}_T|}\sum_{\bm{\sigma}\in\mathcal{D}_T}\mathcal{O}(\bm{\sigma}),
\end{equation}
with that computed on the spin samples produced by the machine $\mathcal{M}_\tau$. After training, we can initialize this machine 
with a random configuration and perform block Gibbs sampling until equilibration. We can then build another spin data set  $\mathcal{S}_
\tau$ with these visible samples and compute the average as,
\begin{equation}
\langle\mathcal{O}(\tau)\rangle_{\mathcal{S}} = \frac{1}{Z_{\bm{\lambda},\tau}}\sum_{\bm{\sigma}}\mathcal{O}(\bm{\sigma})\,\text{e}^{-\mathcal{E}_{\bm{\lambda},\tau}(\bm{\sigma})}\simeq\frac{1}{|\mathcal{S}_\tau|}\sum_{\bm{\sigma}\in\mathcal{S}_\tau}\mathcal{O}(\bm{\sigma}).
\end{equation}
If the machine is properly trained we expect the deviations $\delta\mathcal{O}=|\langle \mathcal{O}(T)\rangle_{\mathcal{D}}-\langle\mathcal{O}
(\tau)\rangle_{\mathcal{S}}|$ to be small. In Fig.~\ref{observables} we plot the energy per spin $E$, the magnetization $M=\langle\sum_i\,
\sigma_i\rangle /N$,  the specific heat $C_V=(\langle E^2\rangle-\langle E\rangle^2) / (NT^2)$ and the magnetic 
susceptiblity $\chi=(\langle M^2\rangle-\langle M\rangle^2) / (NT)$. For the magnetization, we find that even with a number 
of hidden nodes as low as  two (not shown), the machine is able to reproduce the exact behaviour within statistical error.  
This can be explained, since the learning is based on real-space spin configuration samples, and thus the magnetization is 
implicitly encoded into the data set. In the case of the energy however, even though we are computing its value using Eq.~\eqref{IsingHAM}
 applied to the visible units, information about the local energy constraints is not included in the data set.  
This results in a larger discrepancy between the physical value and that generated with the Boltzmann machine.

Most interestingly, it appears that for a given physical system size $N$, the Boltzmann machine with a fixed $n_h$
 learns best away from criticality.
In Fig.~\ref{Cv_scaling}a we plot the scaling of the specific heat 
with the number of hidden nodes in the machine for five different temperatures. When the system is in an ordered or a 
disordered state, the machines trained on the spins of the corresponding data sets are able to reproduce the exact values
within statistical error, irrespective to $n_h$. This is consistent with the weight histograms in Fig.~\ref{histogram}. At high temperature this 
follows from the two layers being quasi decoupled. For low temperatures we have seen that only the hidden nodes that connect 
to positive 
weights (or negative weights, depending on the polarization of the visible layer) are set to 1; increasing the number of hidden nodes will not
affect the activation of the visible units.
Finally, when the system is at criticality, it is still possible to obtain an accurate approximation of the physical distribution, 
however a clear dependency on the finite number of hidden units appears.
As illustrated in Fig.~\ref{Cv_scaling}a, 
in order to converge the specific heat at the critical point, the required $n_h$ is significantly larger than 
for $T$ far above or below the transition. We also note that the same scaling plot for the magnetization (not reported here) shows no clear dependencies on $n_h$. 
Finally, we show in Fig.~\ref{Cv_scaling}b the scaling curves at criticality for different system sizes.  As expected, the threshold in the number of hidden units required for faithful learning of the specific heat grows with increasing $N$.

\begin{figure}[t]
\centering
\includegraphics[width=80mm]{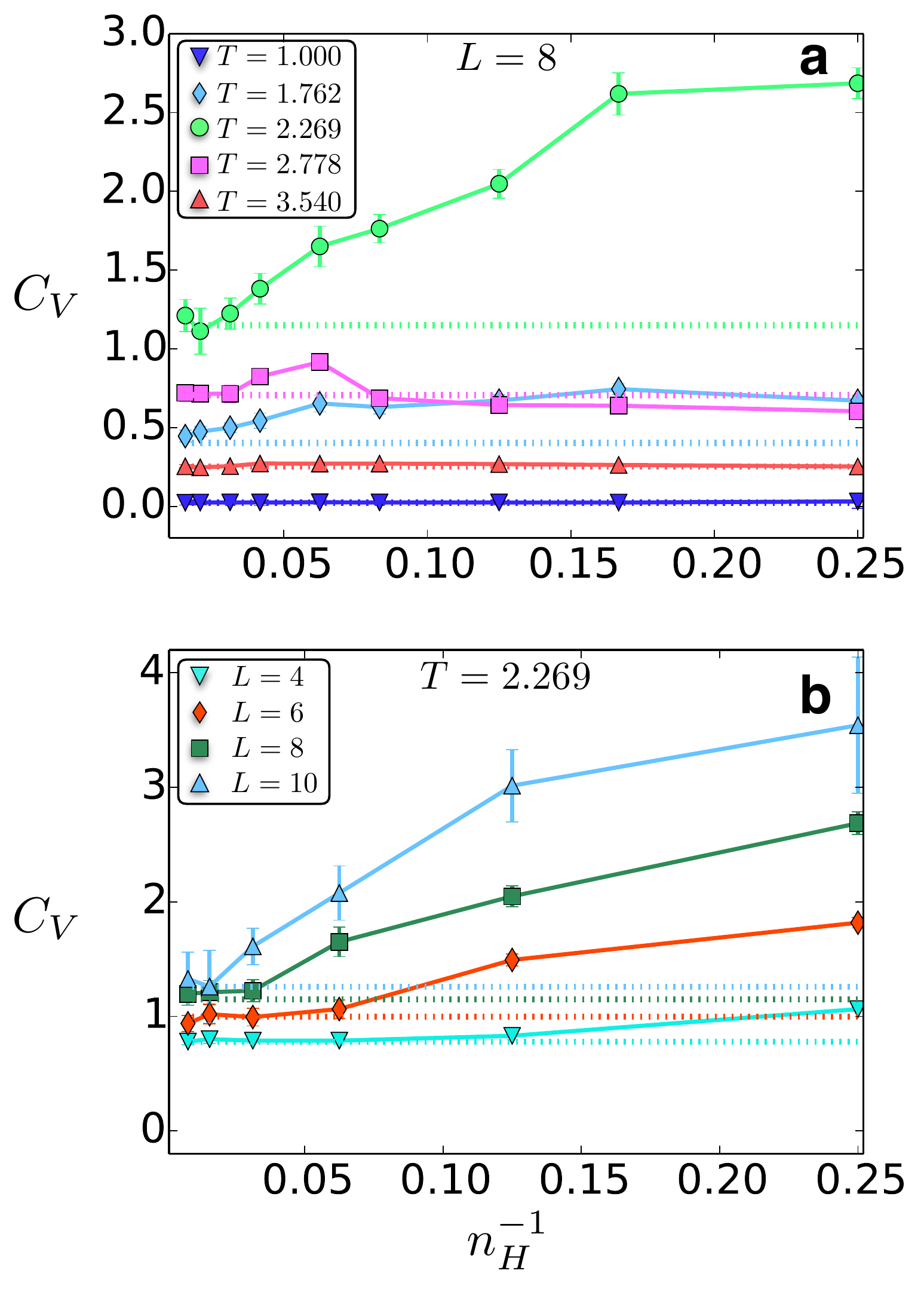}
\caption{Scaling of the specific heat $C_V$ with the number of hidden nodes $n_H$. In ({\bf a}) we show scaling at different temperatures $T$, when the system is ordered (blue and cyan), disordered (red and pink) and critical (green). In ({\bf b}) we show the scaling at criticality for different systems sizes $L$.  Dotted lines represent the exact value computed on the spin configurations of the training dataset.}
\label{Cv_scaling}
\end{figure}

 \section{Conclusions}
We have trained a generative neural network called a restricted Boltzmann machine
to produce a stochastic model of a thermodynamic probability distribution.
The physical distributions were produced by Monte Carlo importance-sampling the
spin configurations of a $d$-dimensional Ising system at different temperatures.
For a small system in $d=1$, we confirm through an exact calculation
that the Boltzmann machine converges to the physical probability distribution 
with sufficient training steps.

For the more difficult problem of the Ising model in $d=2$, where exact
calculations are impossible, we compare thermodynamic observables produced by the 
Boltzmann machine to those calculated directly by Monte Carlo.
Spin samples produced by Monte Carlo were
used to train different machines at distinct temperatures
above, below, and at Ising criticality $T_c$.
Once trained, we evaluated different thermodynamic estimators on the samples 
generated by the Boltzmann machines and show that they faithfully reproduce those calculated directly from 
the Monte Carlo samples. 
For all training instances we fixed the values of the hyper-parameters, and varied the number of hidden nodes. We showed that for $T>T_c$ and $T<T_c$, the Boltzmann machine is able to capture the thermodynamics
with only a few hidden nodes.
However, near $T=T_c$,
the number of hidden nodes required to reproduce 
the specific heat becomes large, reflecting
the increase of fluctuations at criticality.
This growth of hidden nodes required at criticality is reminiscent of the 
connection between deep learning and the renormalization group suggested previously.~\cite{Mehta14} 

Our results demonstrate that Boltzmann machines may serve as a basic research tool for condensed matter and statistical
mechanics, when coupled together with standard Monte Carlo sampling techniques.  One application may be to use the
approximate configurations produced by the trained machine to calculate thermodynamic estimators that may
have been overlooked during the original Monte Carlo sampling (since such configurations are typically discarded).
Similarly, estimator calculation could be completely transferred to the machine, in order to re-distribute these
tasks away from the Monte Carlo procedure.
Conversely, we have demonstrated that the performance of a Boltzmann machine may be evaluated using 
a comparison of thermodynamic observables calculated from both the physical and modelled distribution.  The
conceptual elimination of reliance on the KL divergence may suggest alternatives to evaluating the performance 
of such machines in other applications. 

Among the many possible future applications, 
it would be particularly interesting to train a Boltzmann machine on configurations produced in various bases by quantum Monte Carlo.\cite{Carleo16}
One may ask if a standard restricted machine like studied in the present paper is sufficient to capture quantum correlations,
or if a quantum version of the machine is required.\cite{QBM}  It would also be interesting to understand the relationship between 
the sign problem in calculations of estimators directly in quantum Monte Carlo versus their approximation by 
suitably-trained Boltzmann machines.

\acknowledgments
We would like to thank M. Amin, E. Andriyash, G. Carleo, J. Carrasquilla, B. Kulchytskyy, D. Schwab, and M. Stoudenmire
for many useful discussions.  This research was supported 
by Natural Sciences and Engineering Research Council of Canada, the Canada Research Chair program, the Ontario Trillium Foundation, and the Perimeter Institute for Theoretical Physics.  Simulations were performed on resources provided by the Shared Hierarchical Academic Research Computing 
Network (SHARCNET). Research at Perimeter Institute is supported through Industry Canada and by the 
Province of Ontario through the Ministry of Research \& Innovation.

\bibliography{mybib}
\end{document}